\documentclass[useAMS,usenatbib]{mn2e}
\usepackage{amssymb}
\usepackage{amsmath}
\usepackage{txfonts}
\usepackage{graphicx}
\usepackage{natbib}
\usepackage{rotating}
\usepackage{url}
 
\def\bi {\begin{itemize}}
\def\ei {\end{itemize}}

\def\asca{\textit {ASCA}}

\def\xmm{\textit {XMM-Newton}}
\def\chandra{\textit {Chandra}}
\def\swift{\textit {Swift}}
\def\suzaku{\textit {Suzaku}}

\def\xmmsp{\textit {XMM-Newton }}
\def\psrb{PSR~B1259--63} 
\def\psrbsp{PSR~B1259--63 } 

\def\sax{\textit{Beppo}SAX}

\def\deg {$^\circ$}
\def\beq {\begin{equation}}
\def\eeq {\end{equation}}

\def\lsi {LSI~+61\degr~303}

\usepackage{color}
\definecolor{red}{rgb}{0.7,0,0}
\definecolor{blue}{rgb}{0,0,0.7}

\begin{document}

\title{X-ray  observations of \psrb{} near the 2007 periastron passage}

 \author[M. Chernyakova et al. ]{M. Chernyakova$^{1}$\thanks{E-mail:masha@cp.dias.ie},
 A. Neronov$^{2,3}$, F. Aharonian$^{1,4}$, Y. Uchiyama $^{5}$, T. Takahashi  $^{6,7}$\\
$^{1}$DIAS, 31 Fitzwilliam Place, Dublin 2, Ireland\\
$^{2}$ISDC, Chemin d'Ecogia 16,CH-1290 Versoix, Switzerland \\
$^{3}$Geneva observatory,Ch. des Maillettes 51, 1290 Sauverny, Switzerland \\
$^{4}$MPIKP, PO Box 103980, 69029 Heidelberg, Germany\\
$^{5}$SLAC National Accelerator Laboratory, 2575 Sand Hill Road M/S 29,
Menlo Park, CA 94025, USA.\\
$^{6}$Institute of Space and Astronautical Science/JAXA, Sagamihara, Kanagawa
229-8510, Japan.\\
$^{7}$Department of Physics, University of Tokyo, 7-3-1 Hongo, Bunkyoku,
Tokyo 113-0033, Japan.\\
}
\date{Received $<$date$>$  ; in original form  $<$date$>$ }
\pagerange{\pageref{firstpage}--\pageref{lastpage}} \pubyear{2009}

\maketitle
\label{firstpage}

\begin{abstract}
\psrb\ is  a 48 ms radio pulsar in a highly eccentric 3.4 year orbit with a Be star SS 2883.
Unpulsed $\gamma$-ray, X-ray and radio emission components are observed from the binary system. It is likely that the collision of the pulsar wind with the anisotropic wind of the Be star plays a crucial role in the generation of the observed non-thermal emission.  The 2007 periastron passage  was observed in unprecedented details with \suzaku, \swift, \xmm\ and \chandra\ missions. We present here the results of this campaign and compare them with previous observations. With these data we are able, for the first time, to study the details of the spectral evolution of the source over a 2 months period of the passage of the pulsar close to the Be star. 
New data confirm the pre-periastron spectral hardening,  with the photon index reaching a value smaller than  1.5, observed  during a local flux minimum. If the observed X-ray emission is due to the inverse Compton (IC) losses of the 10 MeV electrons, then such a hard spectrum can be a result of Coulomb losses, or can be related to the existence of the low-energy cut-off in the electron spectrum. Alternatively, if the X-ray emission is a synchrotron emission of very high energy electrons,  the observed hard spectrum can be explained if the high energy electrons are cooled by IC emission in Klein-Nishina regime.
Unfortunately the lack of simultaneous data in the TeV energy band prevents us from making a definite conclusion on the nature of the observed spectral hardening and, therefore, on the origin of the X-ray emission.  
\end{abstract}

\begin{keywords}
{pulsars : individual:   \psrb~ --
 X-rays: binaries -- X-rays: individual:   \psrb~}
\end{keywords}

\section{Introduction}
\psrb{} is a $\sim$48 ms radio pulsar located in an  eccentric (e$\sim$0.87),
3.4 year orbit with a Be star SS 2883 \citep{johnston92}. This system is known to be 
highly variable on an orbital time scale in radio (\citet{johnston05} and references therein),
X-ray (\citet{chernyakova06} and  references therein),  and TeV \citep{aharon05} energy ranges. 
The orbital multi-wavelength variability pattern is determined by the details of the interaction of a relativistic pulsar wind with a strongly anisotropic wind of the companion Be star, composed of a fast, rarefied polar wind and a slow, dense equatorial  decretion disk. The disk of the Be star in the \psrb\ system
is believed to be  tilted with respect to the orbital plane. While the inclination of the disk is not constrained, the line of intersection of the disk plane and
the orbital plane is known to be oriented at  about $90$\degr\ with respect to the major axis of the binary orbit  \citep{wex98,wang04} and the pulsar passes through the disk twice  per orbit.

The unpulsed radio emission from the system appears approximately 
at the moment of the pre-periastron entrance of the pulsar into the equatorial Be star disk. Within several
 days unpulsed radio emission sharply rises to a peak, and then slightly decreases, 
 as the pulsar passes through periastron. The second peak is reached during the second, post-periastron  disk crossing 
 \citep{johnston99,johnston05,connors02}.
 
 
 \asca\ observations of the \psrb\ system in 1994 and 1995 have shown that, similar to the radio light curve, the X-ray light curve has two peaks around the periastron \citep{kaspi95,hirayama99}. The \xmm\ observations of the source's orbital X-ray light curve, with a detailed monitoring of the period when pulsar approaches and  enters the dense equatorial wind of the Be-star prior to periastron,  have shown that the appearance of the unpulsed radio emission is also accompanied by a sharp rise of the X-ray flux  \citep{chernyakova06}. The source spectral evolution has revealed an unexpected hardening of the source spectrum with the smallest value of the photon index $\Gamma\sim 1.2$, and a subsequent softening on the day scale as the pulsar moves deeper  inside the disk. Unfortunately, in 2004 the source became invisible for \xmm\ just after the entrance to the disk, so that the behaviour of the source within the disk and during the second disk crossing remained unclear. In order to clarify this behaviour, we  organised  an intensive X-ray monitoring campaign during the 2007 periastron passage. We have monitored the source with the \suzaku, \xmm, \chandra\ and \swift\ satellites during a $ 2$~months period, which covers both the pre- and post-periastron disk crossings. The results of this campaign are discussed below.
 
This paper is organized as follows: in Section 2 we describe the
details of the  data analysis.  The results are
presented in Section 3, and discussed in Section 4.

\section[]{Observations and Data Analysis}
In 2007 we were able to organize an intensive X-ray monitoring campaign of the \psrb\
system with \suzaku, \xmm, \chandra, and \swift.  The list of observations is given in  Table \ref{data}, where  $\tau$
denotes the number of days after the  periastron passage (27 July 2007,
MJD 54308) and 
 $\phi$ is the true anomaly of the source.

\subsection{\xmm\ observations}
\begin{table}
\caption{Journal of 2007  of \psrb. \label{data}}
\begin{tabular}{ccc@{\,}c@{\,}c@{\,}c@{\,}c@{\,}c}
\hline
\multicolumn{8}{|c|}{\xmm\ observations}\\
\hline
Data& Date & MJD& $\tau$&$\phi$&Exposure&f\_mos1&f\_mos2 \\
 Set&      &    &  (days)&(deg)&(ks)&&         \\
\hline
X11&  2007-07-08 & 54289 &-19 & 86.14&  9.34& $ 1.00^{+0.01}_{-0.01}$&$ 1.03^{+0.01}_{-0.01}$ \\
X12&  2007-07-16 & 54297 &-11 &112.26& 36.54& $ 0.98^{+0.01}_{-0.01}$&$ 1.05^{+0.01}_{-0.01}$ \\
X13&  2007-08-17 & 54329 &+21 &278.57&  6.35& $ 1.01^{+0.01}_{-0.01}$&$ 1.05^{+0.01}_{-0.01}$ \\
\hline
\multicolumn{8}{|c|}{\chandra\ observations}\\
\hline
Ch1&  2007-07-28 & 54309 &1  &187.61&  4.68 \\
Ch2&  2007-08-06 & 54318 &10 &248.87&  4.67 \\
Ch3&  2007-08-24 & 54337 &29 &289.56&  3.15 \\
Ch4&  2007-09-18 & 54362 &54 &308.83&  7.12 \\
\hline
\multicolumn{8}{|c|}{\swift\ observations}\\
\hline
Sw1&  2007-07-07 & 54288.6 &-20 &84.1&  2.72 \\
Sw2&  2007-07-09 & 54290.6 &-18 &88.3&  5.13 \\
Sw3&  2007-07-11 & 54292.3 &-16 &92.6&  4.56 \\
Sw4&  2007-07-13 & 54294.7 &-14 &99.5&  4.34 \\
\hline
\multicolumn{8}{|c|}{\suzaku\ observations}\\
\hline
Data& Date & MJD& $\tau$&$\phi$&Exposure \\
 Set&      &    &  (days)&(deg)&XIS(ks)       \\
\hline
Sz1&2007-07-07&54288.6  &-19.3&84.6 &21.9\\
Sz2&2007-07-09&54290.7  &-17.2&88.1 &19.5\\
Sz3&2007-07-11&54292.6  &-15.3&95.8 &22.7\\
Sz4&2007-07-13&54294.7  &-13.2&102.2&22.9\\
Sz5&2007-07-23&54304.3  &-3.6& 149.6&19.7\\
Sz6&2007-08-03&54315.3	&7.4&  230.0&24.0\\
Sz7&2007-08-18&54330.1	&22.2& 279.6&20.5\\
Sz8&2007-09-05&54348.2	&40.3& 300.4&18.3\\
\hline

\end{tabular}
\end{table}

\begin{figure}
\begin{center}
\includegraphics[width=8cm,angle=0]{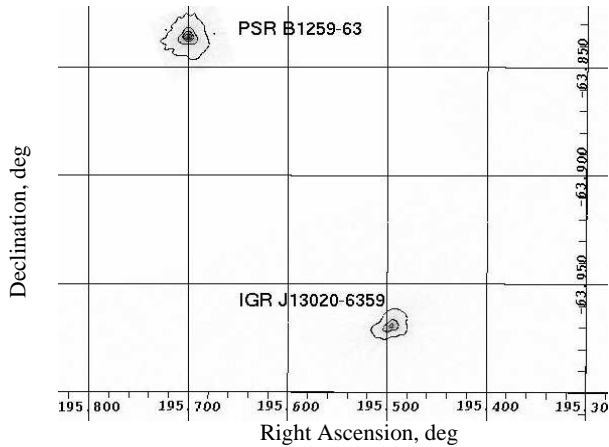}
\end{center}
\caption{Contour plot of the MOS2 \xmm\ field of view for the X12 observation in
equatorial J2000 coordinates. A
total of 4 contours were used with a square root scale between 
5 counts per pixel (outer contour) and 500 counts per pixel (inner
contour).}
\label{xmm_ima}
\end{figure}

In all \xmm\ observations the source was observed with the European Photon Imaging Cameras (EPIC)  MOS1, MOS2 \citep{DenHerder01} and PN \citep{struder01} detectors 
in  the small window mode with a medium filter.
The \xmmsp Observation Data Files (ODFs) were obtained from the
online Science
Archive\footnote{http://xmm.vilspa.esa.es/external/xmm\_data\_acc/xsa/index.shtml} and 
analyzed with the Science Analysis Software ({\sc sas}) v7.1.2. During the
data cleaning, all time intervals in which the count rate in the energy band above 10 keV was higher than 1 cts/sec for the PN detector and/or
higher than 0.35 cts/sec for the MOS detectors, have been removed. In addition,  we discarded the first kilosecond of X12 
and the last two kiloseconds of X13 observations, as these were affected by soft proton flares. 

The event lists for the spectral analysis were extracted from a
15$^\prime$$^\prime$ radius circle at the source position for the X11 MOS1 observation, 
and from a 22.5$^\prime$$^\prime$ radius circle for all other MOS and PN observations. 
We have performed spectral analysis by simultaneously fitting the data of MOS1,MOS2 and PN instruments, leaving the 
inter-calibration factors between the instruments free. The values of the MOS1 and MOS2 inter-calibration factors relative to the PN are
 given in last two columns of Table \ref{data}.

\begin{table}
  \begin{center}
    \caption{Spectral parameters of IGR J13020-6359  during 2007 \swift\  and \xmm\ observations. \label{rxp}}
\begin{tabular}{cccccccc}
\hline
Set&$F$(2-10 keV)                    &$\Gamma$&N$_H^{*}$&$\chi^2$(dof) \\ 
        &$10^{-11}$ ergs cm$^{-2}$s$^{-1}$&        &10$^{22}$cm$^{-2}$ &\\
\hline
Sw1&$ 2.10^{+0.65}_{-0.38}$&$ 1.00^{+1.18}_{-0.14}$&$ 3.00$& 21.969 (11) \\
X11&$ 2.21^{+0.11}_{-0.19}$&$ 0.97^{0.08}_{-0.06}$&$ 2.27^{+0.18}_{-0.12}$& 171.08 (153) \\
Sw2&$ 2.33^{+0.87}_{-0.62}$&$ 1.21^{+0.20}_{-0.19}$&$ 2.56^{+0.46}_{-0.40}$& 12.28 (22) \\
Sw3&$ 1.90^{+1.06}_{-0.64}$&$ 1.44^{+0.27}_{-0.25}$&$ 3.14^{+0.69}_{-0.57}$& 7.28 (14) \\
Sw4&$ 1.83^{+1.53}_{-0.77}$&$ 1.67^{+0.35}_{-0.32}$&$ 3.96^{+1.15}_{-0.93}$& 11.21 (11) \\
X12&$ 3.4^{+0.06}_{-0.09}$&$  0.89^{+0.03}_{-0.02}$&$ 2.33^{+0.07}_{-0.05}$& 534.78 (422) \\
X13&$ 2.89^{+0.10}_{-0.22}$&$ 0.87^{+0.07}_{-0.07}$&$ 2.32^{+0.15}_{-0.14}$& 151.03 (150) \\
\hline 
\end{tabular}
\end{center}
\begin{flushleft} 
$^*$ Due to the lack of statistics we fixed the value of N$_H$ in the Sw1 observation. 
 
\end{flushleft} 
\end{table} 
 

\begin{figure*}
\begin{center}
\includegraphics[width=15cm,angle=0]{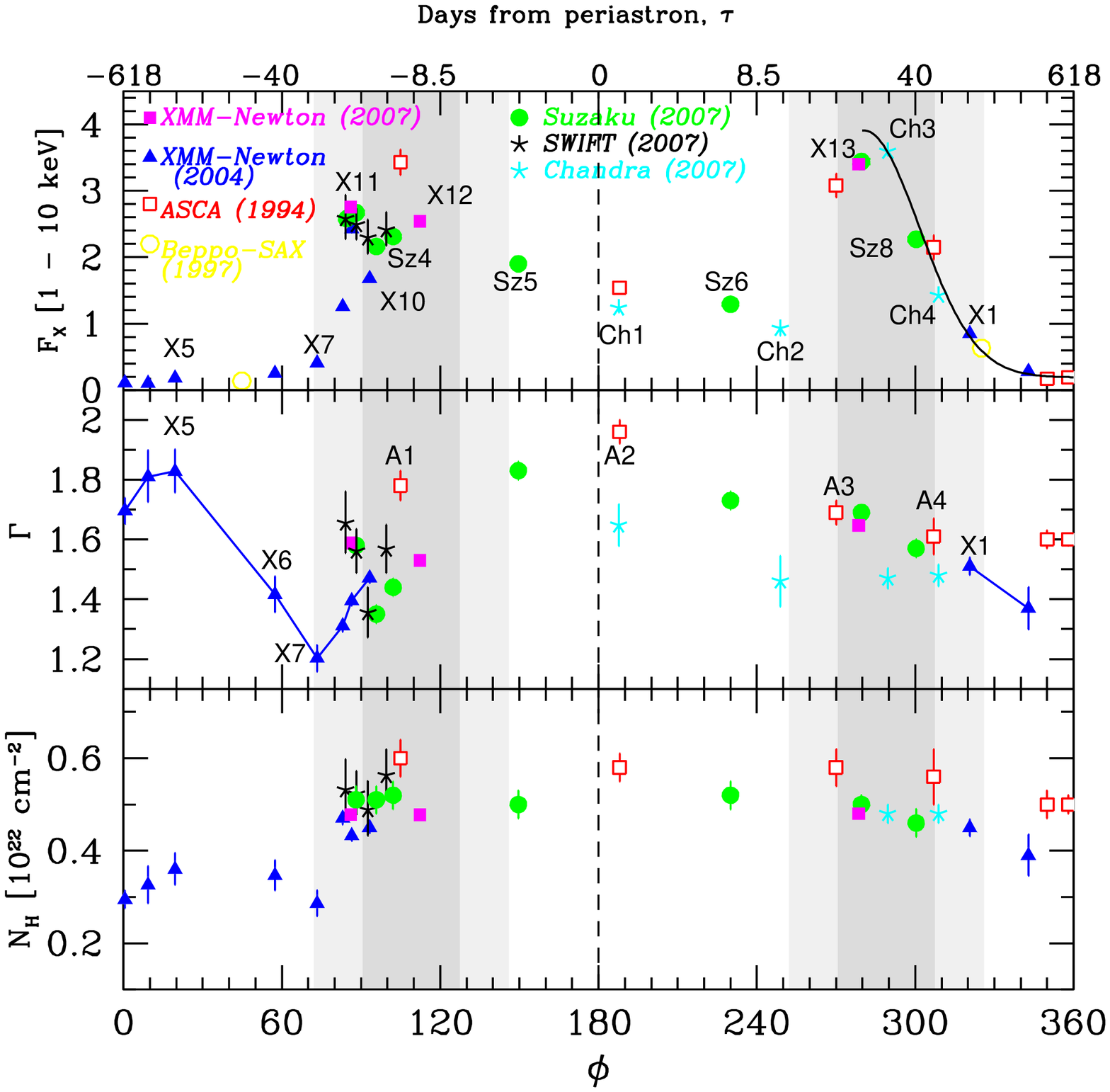}
\end{center}
\caption{\psrb{} orbital  evolution  of  (1-10 keV) light curve (top panel),spectral index (middle panel) and hydrogen column density (bottom panel), as seen with 
Suzaku,\xmm,\swift\ and \chandra\ during the 2007 periastron passage along with 
the old \xmm,\sax\,  and \asca\ observations. 1 -- 10 keV flux of the source is given in units of $10^{-11}$ ergs cm$^{-2}$s$^{-1}$. 
Shaded area corresponds to the disk position proposed in Chernyakova et al. (2006).
The solid line on the top panel is a fit with a Gaussian decay model (see text). To guide the eye we have connected old \xmm\ data on the middle panel. 
}
\label{spechist}
\end{figure*}

\begin{figure*}
\begin{center}
\includegraphics[width=15cm,angle=0]{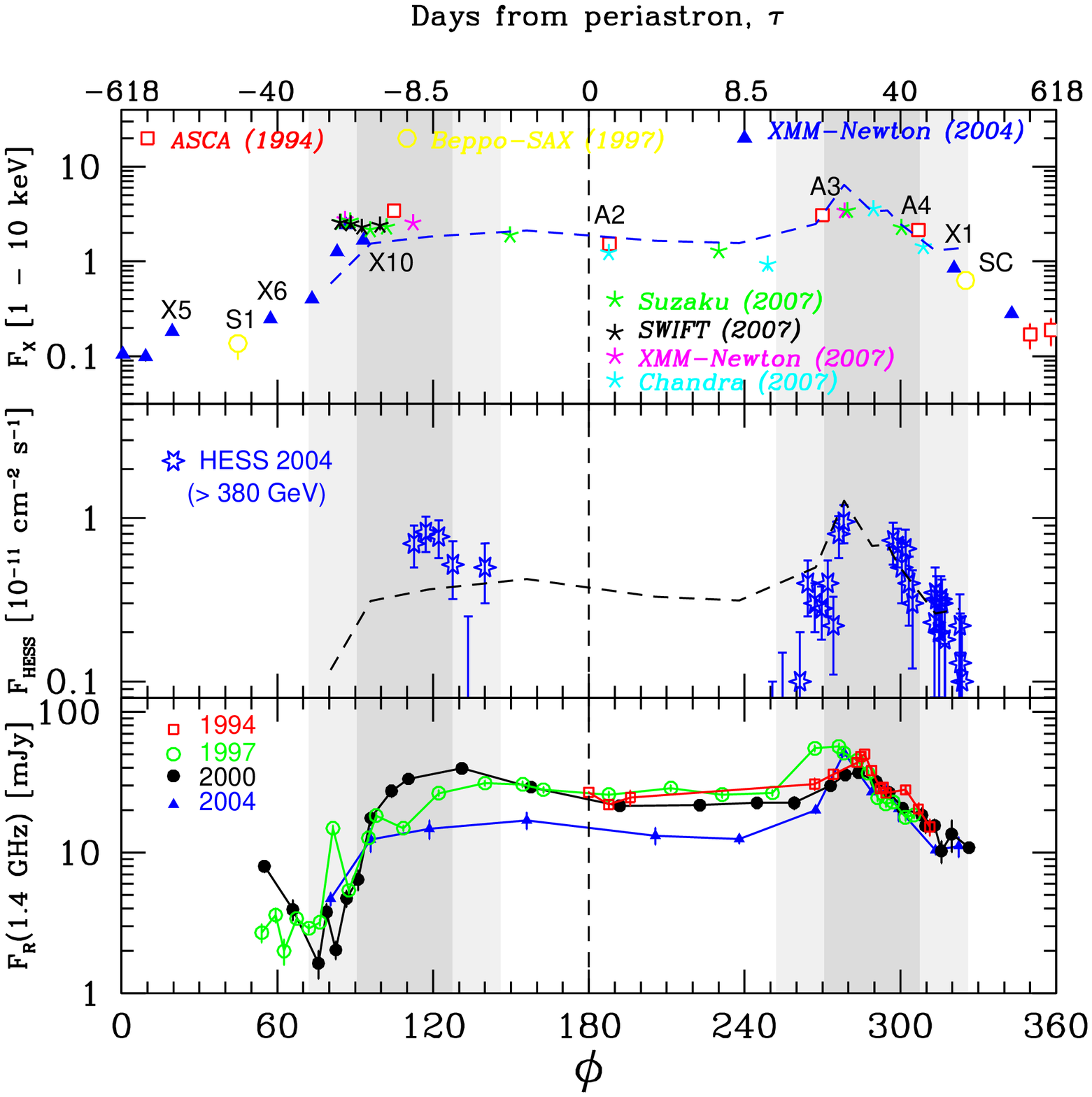}
\end{center}
\caption{\textit{Top panel:}\psrb{} orbital light curve (1-10 keV), as seen with 
Suzaku,\xmm,\swift\ and \chandra\ during the 2007 periastron passage along with 
the old \xmm,\sax\,  and \asca\ observations.  1 -- 10 keV flux of the source is given in units of $10^{-11}$ ergs cm$^{-2}$s$^{-1}$.\textit{middle panel:} 2004 HESS orbital light curve.
\textit{bottom panel:} Collection of historical 1.4 GHz radio light curve during different periastron
passages. In order to compare the orbital behaviour at different wavelength we show with  a black dash line the scaled 2004 radio curve on top and middle panels.
Shaded area corresponds to the disk position proposed in Chernyakova et al. (2006).  
}
\label{3lc}
\end{figure*}


\subsection{\chandra\ observations}

During our 2007 monitoring campaign,  \psrb\ was observed with \chandra\ \citep{weisskopf00} four times, see
Table \ref{data}. The first two observations (Ch1 and Ch2) were done in the Faint mode with HETG grating.
In order to reduce the pile-up, only 1/8 of the S3 detector was in use, which reduces the nominal frame integration 
time from 3.2 seconds to  0.4 seconds. The observations 
Ch3 and Ch4  were done
in CC33\_FAINT mode, because of the higher expected source flux level. This allowed us to obtain good quality results not affected by a pile-up.
Similar to the first two observations, only the S3 detector was in use. 
In the data analysis we used the Chandra Interactive Analysis of Observations software
package (CIAO ver. 3.4) and the CALDB version 3.4.0.
 The tool \texttt{celldetect}  found no 
sources other than \psrb\ in the field of view. We have used the generated
region file for the spectral extraction, and collected background events from a nearby region
of the same form and size. The resulting 
spectra were grouped  to have at least 
30 counts per energy bin. 
\psrb\ showed a moderate level of activity, with an average 
count rate of 0.21 cts/s in Ch1 observations and  0.15 cts/s in Ch2 observation, which corresponds
to a pile-up affection at the 15\%  and 10\% levels, respectively  \footnote{Pile-up affection level
was calculated using PIMMS v3.9d tool, available at http://cxc.harvard.edu/toolkit/pimms.jsp. Using this tool we  assumed that emission can be described as an absorbed power law  with parameters listed in Table \ref{summary}. } It is worth to noteworthy that \cite{kishishita09} in their analysis of Chandra LS 5039 data  found that even a relative small pile-up fraction can result in somewhat harder photon indexes. Thus the absolute numbers found for Ch1 and Ch2 should be treated with some care.

\subsection{\swift\ observations}

The  \swift/XRT \citep{gehrels04} data  were taken in photon mode with a $500\times500$ window
size. We have processed all the data with standard procedures 
using the FTOOLS\footnote{See http://heasarc.gsfc.nasa.gov/docs/software} 
task \texttt{xrtpipeline} (version 0.11.6 under the
 HEAsoft package 6.4.0). 
 We have extracted source events from a circular region with a radius of
 20 pixels (1 pixel $\simeq 2.27''$). To account for the background, 
 we have extracted events from a nearby circle of the same radius. 
 Due to the low count rate (less than 0.4 cts/s) 
 no pile-up correction was needed. The spectral data were rebinned with a minimum of 25 
 counts per energy
 bin for the $\chi^2$ fitting. The ancillary response file was generated 
 with \texttt{xrtmkarf}, taking into account  vignetting, and the
 point-spread function corrections. In our
 analysis we have used the \texttt{swxpc0to12s0\_20010101v010.rmf} spectral 
 redistribution matrix for observations Sw1, Sw2 and Sw3 and the 
 \texttt{swxpc0to12s6\_20010101v010.rmf} spectral redistribution matrix for the observation Sw4.
 
\subsection{Suzaku observations}

Suzaku has intensively monitored the \psrb\ 2007 periastron passage. Observations
were performed with the X-ray Imaging spectrometer (XIS: \cite{koyama07}) and the Hard X-ray detector (HXD: \cite{takahashi07}). These data were first presented in  \cite{uchiyama09}. 
For completeness we
present  the list of \suzaku\ observations in Table \ref{data}.

\section{Results}
\subsection{Imaging Analysis}
X-ray emission from \psrb\ has been clearly detected in all observations with all instruments.  Apart from \psrb, another X-ray binary, IGR J13020-6359 (identified with 2RXP J130159.6-635806 in \cite{cher05}), located 
10 arcminutes away to the north-west, is detected in the MOS2 \xmm\  and XRT \swift\ fields of view. Figure \ref{xmm_ima}  shows the contour
plot of \xmm\ field of view for the X12 observation, in which both sources are visible. IGR J13020-6359 was not detected in MOS1, because  some of its CCDs were turned off during the observation. This source is found to be in a low activity state, with the flux comparable to the flux of \psrb.  Table \ref{rxp} summarizes the spectral parameters of IGR J13020-6359 found in our observations.

\subsection{The X-ray light curve}

The upper panel of Fig. \ref{spechist} shows the X-ray orbital  light curve of
\psrb\ system. The figure summarizes all the available X-ray data on the source  in 1-10 keV energy range. The historical data of  \xmm\ (X1 -- X10) and \sax\ points are taken from \cite{chernyakova06}, \asca\ data are taken from  \cite{hirayama99}.
 Observations made with different instruments at close orbital phases are
consistent with each other, demonstrating satisfactory inter-calibration between different instruments: the
points marking Sz1 and Sz2 and X11 and a \swift\ observation at the phase $\theta\sim 85^\circ$ are superimposed on each other.
Some of our new observations are done at  orbital phases close to the ones of historical \asca\  observations (e.g. Ch1). One can see that the system's orbital light curve does not exhibit strong orbit-to-orbit variations - all fluxes measured by \asca\ are consistent with the ones measured 13 years later. Our new data also confirm a local minimum at the phase  $\theta\simeq 90^\circ$, first observed with \xmm\
during the 2004 periastron passage (X9 and X10 observations).

Stability of the orbital light curve allows us to use old and new data simultaneously while analyzing the 
flux orbital evolution. The first (pre-periastron) entrance of the pulsar into the equatorial disk of the Be star is accompanied by a sharp rise of the X-ray flux  (points X7 - X9),  by  a factor of six in seven days ($\Delta\theta=13$\deg). This period of the sharp rise of the flux is followed by a period of variability, most probably related to the interaction of the pulsar wind with the Be star disk. During this period, the X-ray flux decreases by a factor of 1.5 for two days ($\Delta\theta=7$\deg) and then rises again for the subsequent two days ($\Delta\theta=6$\deg). The decay of the flux after the exit of the pulsar from the disk is slower than the rise at the disk entrance.  The flux decreases by a factor $\sim 2.5$ up to the moment of the post-periastron entrance to the disk.  The second entrance to the disk is again accompanied by  a
 sharp rise of the flux, by a factor of 3.5 in less than ten days ($\Delta\theta<30$\deg). The second peak is followed by an adiabatic decay with the characteristic decay scale of 30\deg, illustrated in Fig. \ref{spechist} by  solid curve, representing the best fit of the data with Gaussian decay model ($f(\theta)=a*e^{-(\theta-\theta_0)^2/\Delta\theta_0^2}+c$, $\theta_0=280$\deg,$\Delta\theta_0=30$\deg).

For comparison we also show in  Fig. \ref{3lc} the TeV light curve
of 2004 HESS observation \citep{aharon05}  and radio
\citep{johnston99,connors02,johnston05} light curves of different
 periastron passages. It is seen that the general flux behaviour (rapid rise at the disk entrance, followed by a slow decay)  in all energy bands is  similar. To make this point more clear we add the scaled radio light curve of 2004 periastron passage to panels showing the X-ray and TeV data (dashed line).

\begin{table} 
  \begin{center}
\caption{Spectral parameters for 2007 observations of \psrb.\label{summary}}  
\begin{tabular}{ccccccc}
\hline
Set&$\phi$& F$^*$(1-10 keV) &$\Gamma$ &N$_H^{**}$&$\chi^2$(ndof)\\ 
\hline
Sw1&84.1&$ 2.58^{+0.36}_{-0.31}$&$ 1.65^{+0.1}_{-0.1}$& $0.53^{+0.07}_{-0.06}$& 33 (35) \\
Sz1 &84.58 &2.58$\pm{0.03}$& $1.64\pm{0.02}$& $0.5\pm{0.02}$& 375(354) \\ 
X11& 86.14&$ 2.76^{+0.04}_{-0.04}$&$ 1.589^{+0.010}_{-0.010}$&$ 0.478^{+0.006}_{-0.006}$& 1490 (1575)\\
Sw2&88.3&$ 2.49^{+0.25}_{-0.22}$&$ 1.56^{+0.07}_{-0.07}$& $0.52^{+0.05}_{-0.05}$& 63 (654) \\
Sz2 &88.10 &2.67$\pm{0.03}$& $1.58\pm{0.03}$& $0.51\pm{0.03}$& 321(309) \\ 
Sw3&92.6&$ 2.39^{+0.27}_{-0.24}$&$ 1.35^{+0.09}_{-0.08}$& $0.49^{+0.06}_{-0.06}$& 41 (48) \\
Sz3 &95.80 &2.16$\pm{0.03}$& $1.35\pm{0.03}$& $0.51\pm{0.03}$& 324(272) \\ 
Sw4&99.5&$ 2.41^{+0.27}_{-0.24}$&$ 1.57^{+0.08}_{-0.08}$& $0.56^{+0.06}_{-0.05}$& 42 (53) \\
Sz4 &102.17&2.31$\pm{0.03}$& $1.44\pm{0.03}$& $0.52\pm{0.03}$& 387(298) \\
X12&112.26&$ 2.54^{+0.02}_{-0.02}$&$ 1.530^{+0.005}_{-0.005}$&$ 0.478^{+0.003}_{-0.003}$& 2792 (2543)\\
Sz5 &149.61&1.90$\pm{0.02}$& $1.83\pm{0.03}$& $0.50\pm{0.03}$&212(249)  \\ 
Ch1&187.61& $ 1.24^{+0.06}_{-0.07}$&$ 1.65^{+0.07}_{-0.07}$&$ 0.48$& 16(30)\\		    
Sz6 &230.02&1.29$\pm{0.02}$& $1.73\pm{0.03}$& $0.52\pm{0.03}$& 189(205) \\
Ch2&248.87& $ 0.93^{+0.06}_{-0.11}$&$ 1.46^{+0.08}_{-0.09}$&$ 0.48$& 23 (22)\\		    
X13&278.57&$ 3.40^{+0.04}_{-0.04}$&$ 1.647^{+0.010}_{-0.009}$&$ 0.480^{+0.005}_{-0.005}$& 1648 (1595)\\
Sz7 &279.61&3.44$\pm{0.04}$& $1.69\pm{0.02}$& $0.50\pm{0.02}$& 406(402) \\ 
Ch3&289.56& $ 3.59^{+0.06}_{-0.07}$&$ 1.47^{+0.03}_{-0.03}$&$ 0.48^{+0.02}_{-0.02}$& 187 (196)\\ 
Sz8 &300.35&2.27$\pm{0.03}$& $1.69\pm{0.02}$& $0.46\pm{0.03}$& 283(253) \\ 
Ch4&308.83& $ 1.43^{+0.03}_{-0.03}$&$ 1.48^{+0.04}_{-0.04}$&$ 0.48^{+0.02}_{-0.02}$& 193 (200)\\ 

\hline
\end{tabular}
\end{center}
\begin{flushleft} 
$^*$ In units of $10^{-11}$ erg cm$^{-2}$ s$^{-1}$.\\ 
$^{**}$ In units of $10^{22}$ cm$^{-2}$. 
 
\end{flushleft} 
\end{table}

\subsection{Spectral Analysis}
\begin{figure}
\begin{center}
\includegraphics[width=8cm,angle=0]{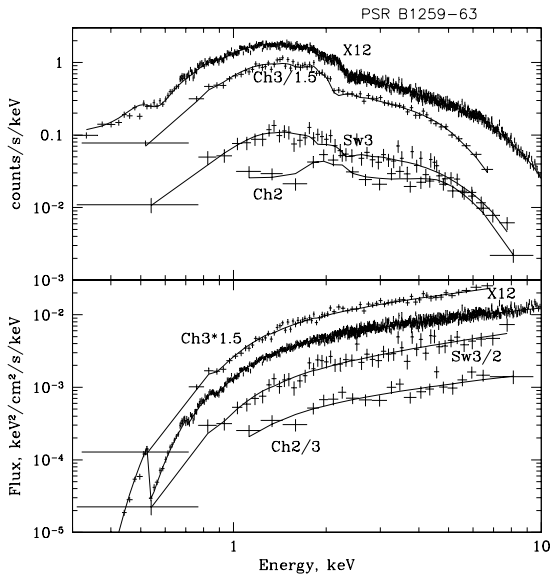}
\end{center}
\caption{ Folded(top panel) and unfolded (bottom panel) \psrbsp spectra during the
 X12 (PN data), Sw3, Ch2 and Ch3 observations. In order to make the figure clearer we have multiplied the observed 
spectra by the specified factors. }
\label{spectry}
\end{figure}

The spectral analysis was done with the NASA/GSFC XSPEC v11.3.2 software package. 
In Fig.~\ref{spectry} the folded and unfolded spectra of \psrb~ for X12, Sw3, Ch2,
and Ch3 observations are shown. In order to make the figure clear, we have multiplied the observed 
spectra by the specified factors.

\begin{figure*}
\begin{center}
\includegraphics[width=5cm,angle=0]{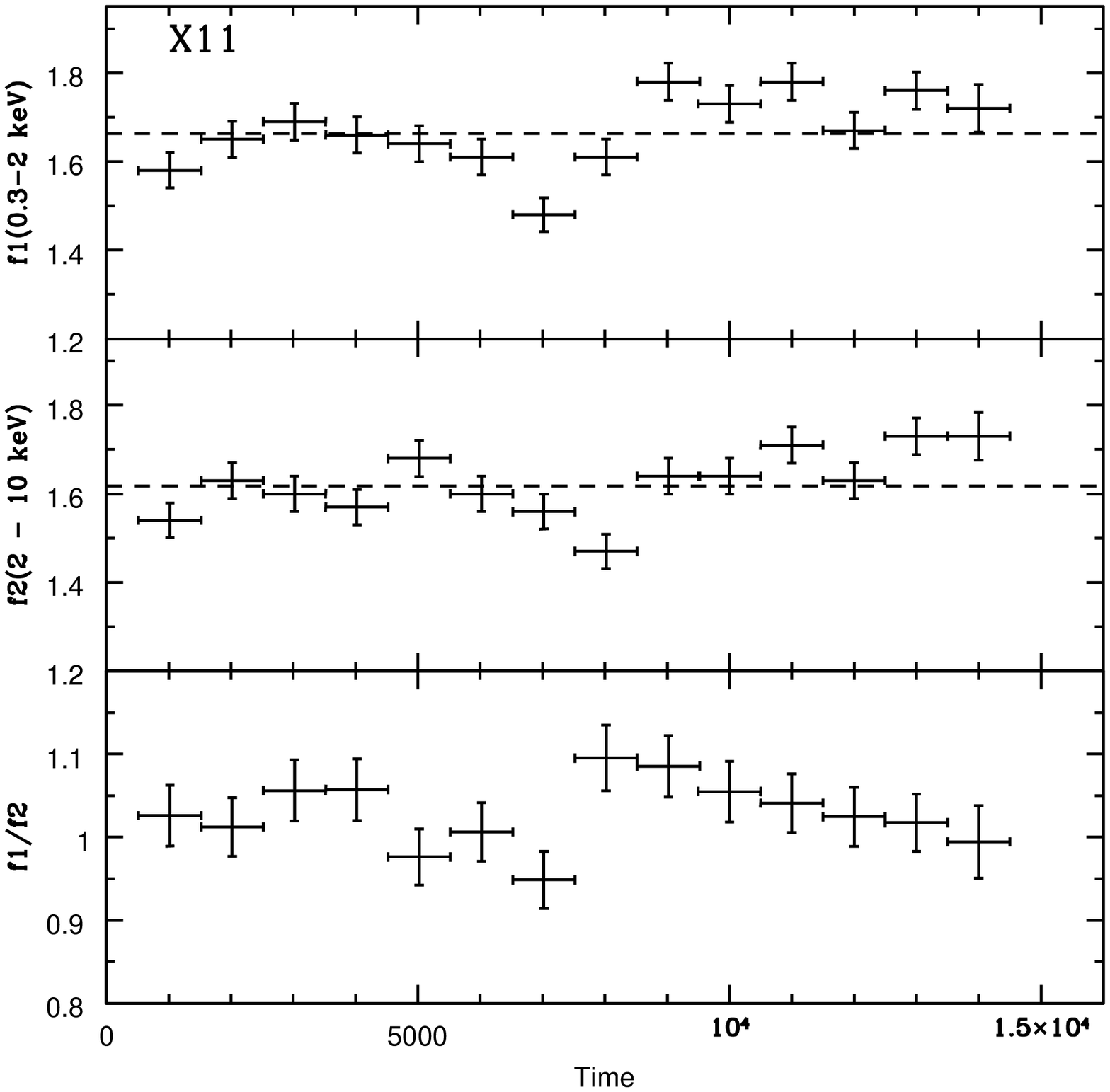}
\includegraphics[width=5cm,angle=0]{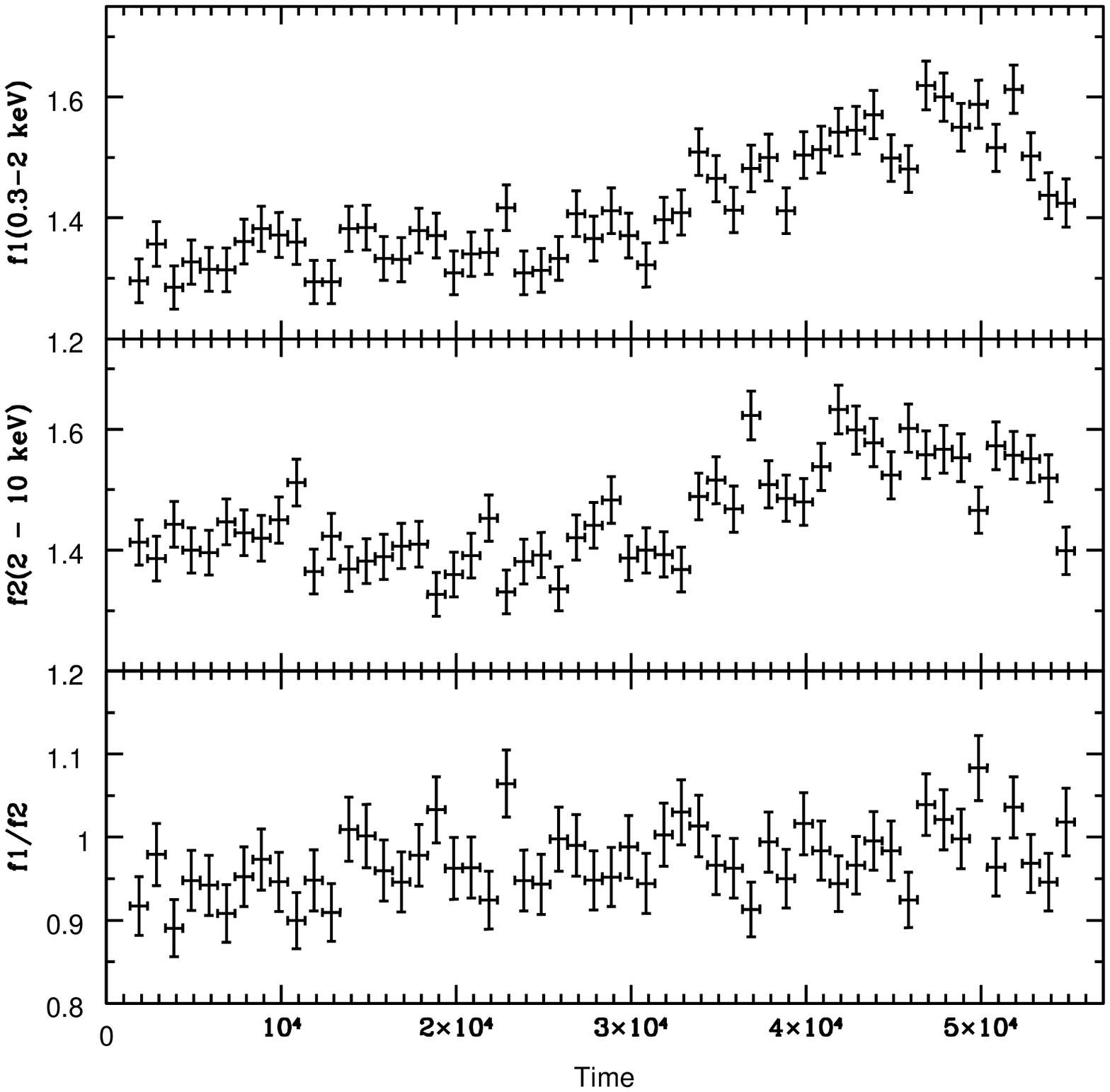}
\includegraphics[width=5cm,angle=0]{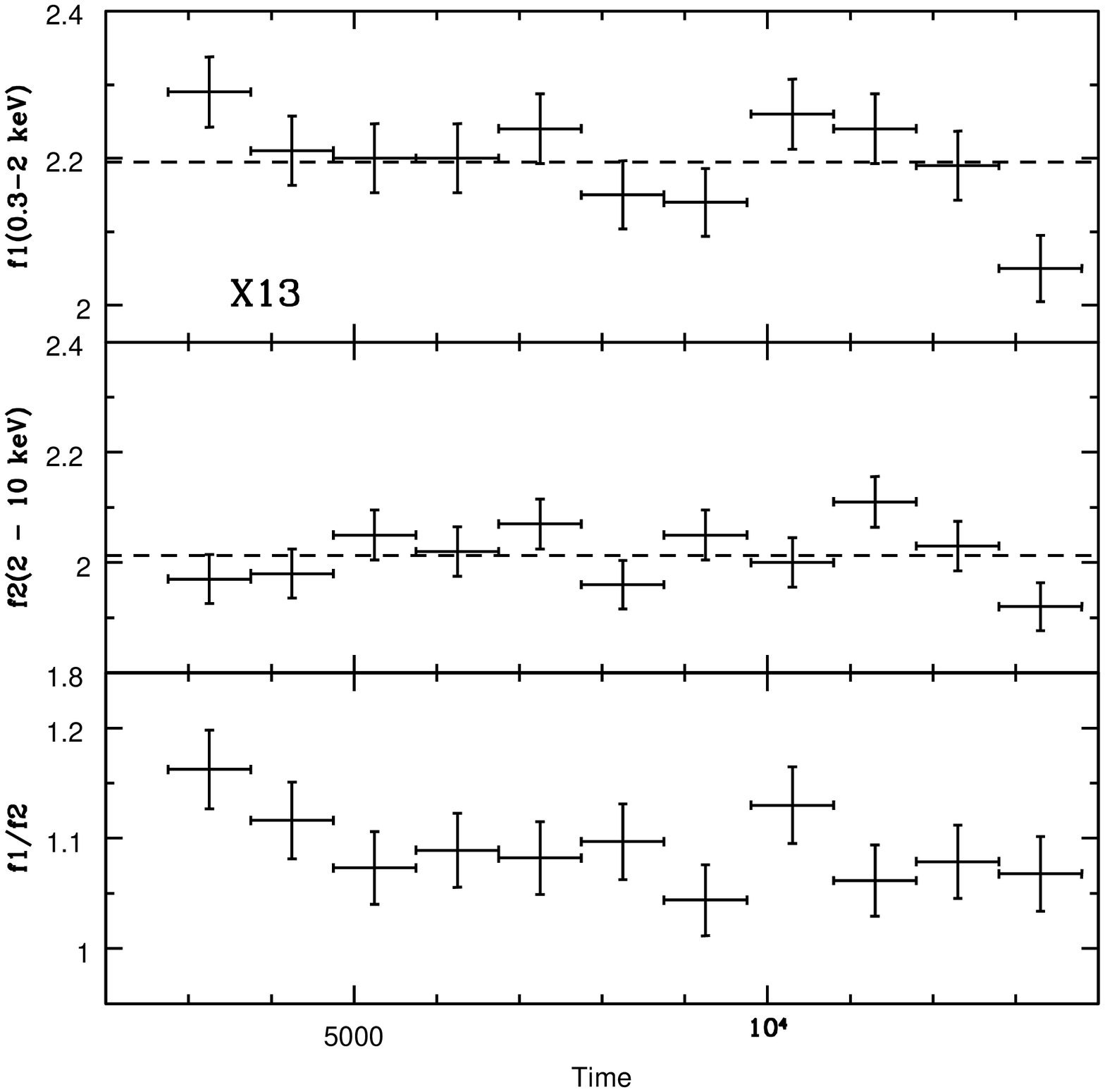}
\end{center}
\caption{Light Curves of X11(left), X12(middle) and x13(right) observations in 0.3 - 2 keV (top panel), 2 - 10 keV (middle panel) along with
hardness ratio (bottom panel). Time bin size is equal to 1ks. Dashed line indicates the best fit of the data with a constant model. }
\label{lcxmm}
\end{figure*}
A simple power law model with  photoelectrical absorption describes the data well in most observations,
with no evidence for any line features. In Table \ref{summary}
we present the results of the three parameter fits to the Suzaku, \xmm, \chandra\ and \swift\ data in
the 0.5 -- 10 keV energy range.
The uncertainties are given at the $1\sigma$
statistical level and do not include systematic uncertainties. The quality of the first two
\chandra\ observations prevent the simultaneous determination of the spectral slope and absorption column density, so we decided to fix the latter to
the value of $N_H=4.8\times10^{21}$cm$^{-1}$, consistent with the value found in \xmm\ observations.
Orbital evolution of the spectral parameters is shown in Figure \ref{spechist}.
The high value of the reduced $\chi^2$ for Sz3 (1.19 for 272 d.o.f.) and Sz4 (1.30 for 298 d.o.f.) observations was studied in details in \cite{uchiyama09}. It was found that for these observations broad band (0.6-50 keV) spectrum is much better fitted with a broken power law model with a spectral break from $\Gamma_1=1.25\pm0.04$ below E$_{br}\sim 5$ keV to $\Gamma_2=1.6\pm0.05$ above.


\subsection{Timing Analysis}

 Within one orbit \psrb\ is known to be variable at different time scales. Apart from the orbital time scale (several years), fast, day-scale, variability of the flux and spectral characteristic is observed during the periastron passage.  Variability at  time scales much shorter than the orbital time scale is observed in other TeV binaries. For example, the X-ray emission from the binary \lsi\,  is characterized by a spectral energy distribution similar to the one of \psrb\ (which might point to the similar mechanism of the X-ray emission)  with variability time scales of down to $\sim 1$~ksec \citep{sidoli06} and possibly even shorter \citep{rxte}. Such short time scale variability could be related to the clumpy structure of the wind of the Be star \citep{zdz08}. If  so, then the clumpy structure of the Be star wind should also lead to the short-time scale variability of emission from the \psrb\ system. In order to study the variability on the short  timescales,  we have analyzed the light curves in the 0.3-2 keV and 2-10 kev energy ranges in
individual X-ray observations.

Figure \ref{lcxmm} shows the soft and hard energy band light curves (1ks
time bins) of the  three 2008 \xmm\ observations of the source. The lower panels of this figure show the 
 hardness ratios. 
 
 The source flux appears to be variable at the observation ($\sim 10$~ksec) time scale at least in the X11 and X12 observations. In the case of the observation X11, the  best fit of the soft (hard) band light curves with a constant flux, shown in the Figure \ref{lcxmm}, gives a $\chi^2=55.88$ ($\chi^2=42.52$) for 13 degrees of freedom. The probability that  the soft (hard) band flux stayed constant over the entire observation is $3\times 10^{-7}$  ($5\times 10^{-5}$), i.e. the constant flux hypothesis is ruled out at a $\sim 6\sigma$  ($\sim 5\sigma$) level.    

To study the details of variability at the $\sim 10$~ks time scale, we  arranged a long $\sim 50$~ks continuous exposure during the observation X12. The flux of the source
was stable (1.4 cts/s in 2-10 keV energy range) during the first $\sim 30$~ks of the observation and then during the following 10 ks grew up 
to a count rate of 1.6 cts/s. No sharp variations of the flux and hardness ratio on the short time
scales have been observed.  No significant variations of the flux are detected in the \xmm\ observation  X13: the scatter of the data points in the light curve shown in the right panel of Fig. \ref{lcxmm} does not exceed the 3$\sigma$ limit in the soft or the 2$\sigma$ limit in the hard energy bands. 

A possible episode of faster ($<10$~ks time scale) flux variations was detected in the X11 observation, which is close to the moment of pulsar entrance into the disk and is in the middle of the period of strong spectral variability. In order to find the minimal variability time scale in this observation we have applied the structure function analysis \citep{simonetti85} to the X-ray light curve.

\begin{figure}
\includegraphics[width=0.9\columnwidth,angle=0]{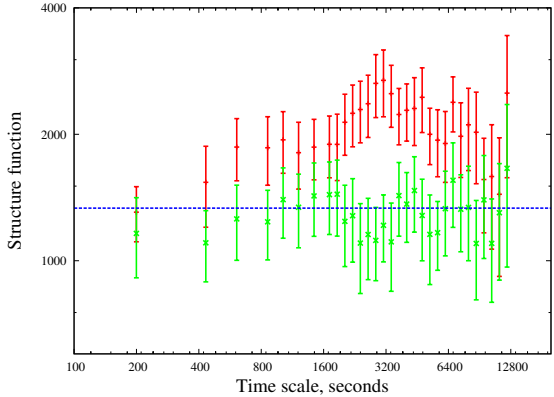}
\caption{Red: structure function of the 0.3-2~keV band light curve of the observation X11. Dashed horizontal line shows the low plateau. Green: example of the structure function of a simulated light curve in which flux variations are due to the statistical scatter of the signal.}
\label{sf}
\end{figure}

Fig. \ref{sf} shows the structure functions calculated for the soft band light curve of X11 observation. The error bars of the structure function are estimated via the simulation of  $10^4$ light curves in which the values of the flux in each time bin are scattered around the measured values of the flux. The structure function deviates by $3\sigma$ from the low plateau value at the time scale $\tau\simeq 3$~ks, which means that the light curve is variable at this time scale. For comparison we show, in the same figure, an example structure function of one of the simulated light curves, in which the flux variations are only due to the statistical uncertainty of the signal. One can see that the structure function of the simulated light curve is consistent with the lower plateau value in all the considered time range.

Evidence of the variability of the source on the several kiloseconds time scale was also reported in some of ROSAT \citep{cominsky94} and \suzaku\ \citep{uchiyama09} observations.

\begin{figure}
\begin{center}
\includegraphics[width=8cm,angle=0]{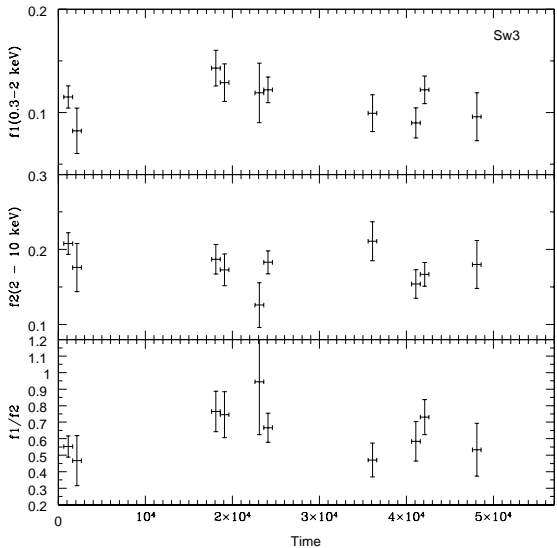}
\end{center}
\caption{Light Curves of Sw3 observation in 0.3 - 2 keV (top panel), 2 - 10 keV (middle panel) along with
hardness ratio (bottom panel). Time bin size is equal to 1ks.}
\label{lcsw3}
\end{figure}

An example of the \swift\ light curve
during the Sw3 observation is shown in Figure \ref{lcsw3}. One can see that the source monitoring strategy was to take a set of $\sim 1$~ks long snap-shots separated by longer (several kiloseconds) time intervals. From Fig. \ref{lcsw3} one can see that no significant variability at the time scale $\sim 10^4$~s is found in the \swift\ data.

\section{Discussion}

The 2007 observational campaign confirms the previous finding \citep{chernyakova06} that during the first, pre-periastron passage of the disk, the X-ray spectrum of the source hardens on a day scale reaching a value of the  photon index smaller than 1.5. During the 2004 periastron passage, a spectral state with photon index  $\Gamma_{\rm ph}\simeq 1.2$ was observed at the beginning of the  rapid growth of the X-ray flux at the disk entrance ($\phi=73$\deg). Remarkable, the time of the subsequent spectral softening coincided with the time of a step-like increase of the hydrogen column density ($\phi=83$\deg), which can be related to the penetration of the pulsar in the dense equatorial disk. The observed hardening, along with the successive softening and sharp flux rise, was attributed by \citet{chernyakova06} to the injection of high energy electrons at the disk entrance (e.g. due to the proton-proton collisions), or to a sharp decrease of the high energy electron's escape velocity accompanied with the modification of their spectrum by  Coulomb losses. An alternative explanation of the hard X-ray spectrum observed at the disk entrance is possible within a model in which X-ray emission is produced via the synchrotron mechanism \citep{khangulyan07, uchiyama09}. In this case hardening of the spectrum to  photon indexes $\Gamma_{\rm ph}<1.5$ on a day time scale can be achieved if the electron energy loss is dominated by  inverse Compton (IC) scattering losses  in the Klein-Nishina regime \citep{khan05}. The Klein-Nishina regime of IC scattering becomes important   at  energies above $E_e\ge m_e^2c^4/(2.7kT_*)\sim 30$~GeV, where $T_*$ is the temperature of the Be star. Decrease of the IC cross-section at high energies leads to the decrease of the efficiency of the IC cooling of electrons and, as a result, to the hardening of the electron spectrum. Electrons with a power law energy distribution $N_e\propto E_e^{-\Gamma} $ cooled in the Klein-Nishina regime form a spectrum  $N_e(E_e)\propto E_e^{-\Gamma+1}\times\left[\ln\frac{4EkT}{m^2c^4}-\frac{2\Gamma}{\Gamma^2-1}-0.6472\right]^{-1}$, much harder than $N_e(E_e)\propto E_e^{-(\Gamma+1)}$ in Thompson limit \citep{bg70}. Thus the resulting   synchrotron spectrum is proportional to  $\epsilon^{-\Gamma/2}$ and can be harder than 1.5. 

The 2007 data did not cover the period of the entrance to the disk, so  it is not possible to check if the spectral hardening preceeding the flux growth repeats from orbit to orbit. Instead, the photon index $\Gamma_{\rm ph}\simeq 1.3$ was observed in 2007 at almost  the maximum of the X-ray flare associated to the pre-periastron disk passage. More precisely, the time of the hardening and subsequent softening of the spectrum in the 2007 data coincides with a local "dip" in the X-ray light curve during the broad flare associated to the disk passage.  No strong variations of the hydrogen column density are noticeable at the moment of the hardening of the spectrum. Detection of the spectral break in the X-ray spectrum of the source at $E\simeq 5$~keV in {\it Suzaku} observation, which reveals the hard state with $\Gamma_{\rm ph}\simeq 1.3$ \citep{uchiyama09}, provides an additional clue for the understanding of the nature of the observed spectral hardening.

These two models of spectral hardening  could be readily distinguished via multi-wavelength observations. In the case of the IC mechanism of the X-ray emission, the energies of electrons responsible for the X-ray emission are in the range of $\sim 10$~MeV, while in the synchrotron model the X-ray emitting electrons have multi-TeV energies. A necessary condition for the hardening of the spectrum beyond $\Gamma=1.5$ in the synchrotron model is that the energy losses of multi-TeV electrons are dominated by the IC loss. This means that the expected TeV-band  IC luminosity of the source  at the moment of the spectral hardening is  larger than the X-ray luminosity.  Unfortunately, the absence of the TeV observations simultaneous with the X-ray observations in 2007 does not allow us to distinguish between the two models.  

If the hard X-ray spectrum can be explained by Coulomb losses of the  $\sim10$~MeV electrons, then  the break at the $\sim5$~keV energy, observed by Suzaku, has to be ascribed to the so-called "Coulomb" break in the electron spectrum. Such a break appears at the energy $E_C$, at which the rate of the Coulomb energy loss is equal to the IC energy loss rate: $E_C\simeq15\left[D/10^{13}\mbox{cm}\right]\left[n_e/10^{8}\mbox{cm}^{-3}\right]^{1/2}\mbox{ MeV}$, where $n_e$ is the stellar wind density at a distance $D$ (10$^{13}$ cm is a characteristic separation of the companions at orbital phases close to periastron). Electrons with the energy $E_C$ produce IC emission in the energy band $
\epsilon_C\simeq4\left[T_*/3\times 10^4\mbox{ K}\right]\left[E_C/10\mbox{ MeV}\right]^{2}\mbox{ keV}$.
Measurement of the break in the X-ray spectrum of the source enables an estimate of the density of the medium
\begin{equation}
n_e(D\simeq 10^{13}\mbox{ cm})\simeq 10^8\mbox{ cm}^{-3}
\end{equation}
This estimate is consistent with the one expected for an equatorial disk with the radial density profile
$n_e(D)=n_0(D/R_\star)^{-3.5}$ if the disk density close to the surface of the star is $n_0\sim 10^{12}$~cm$^{-3}$.

\begin{figure}
\includegraphics[width=\linewidth]{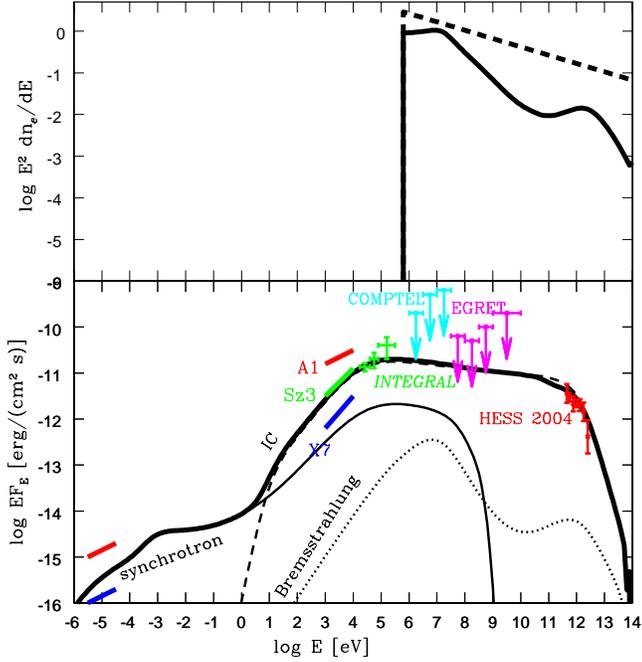}
\caption{Spectral energy distribution of \psrb\  in the model with the power law injection spectrum of electrons with  spectral index $\Gamma_e=2.2$.Top panel shows the initial electron injection spectrum with the dashed line and the spectrum formed in result of cooling and escape of electrons to the distance $D\sim 10^{14}$~cm (solid line). Bottom panel shows the spectra of synchrotron (thin solid line), inverse Compton (dashed thin line) and bremsstrahlung emission (dotted thin line) as well as the overall spectrum modified by the pair production in the photon field of the Be star (thick solid line).Blue and red lines at the radio part of the spectrum  illustrate the slope and possible flux range of the observed radio emission}
\label{fig:IC} 
\end{figure}

If the X-ray emission is produced via the IC mechanism, the break in the X-ray spectrum can be also related to the existence of a low-energy cut-off in the electron spectrum. Indeed, the IC cooling time of the X-ray emitting electrons of the energy $E$ is 
\begin{eqnarray}
\label{tict}
t_{IC(T)}&=&\frac{3\pi m_e^2c^4D^2}{\sigma_TL_\star E}\nonumber\\
&\simeq& 6\times10^5\left[\frac{10^{38}\mbox{erg s}^{-1}}{L_\star}\right]\left[\frac{D}{10^{13}\mbox{cm}}\right]^2\left[\frac{10\mbox{MeV}}{E_e}\right] \mbox s
\end{eqnarray}
where $L_\star$ is the luminosity of the Be star, $m_e$ is the electron mass and $\sigma_T$ is the Thomson cross-section. The IC cooling time is comparable to the escape time  $t_{\rm esc}\simeq D/V\simeq 10^6\left[D/10^{13}\mbox{ cm}\right]\left[V/10^7\mbox{ cm/s}\right]\mbox{ s}$ if the escape velocity is $V\sim 10^7$~cm/s. If the  IC emitting electrons are initially injected at energies much larger than 10~MeV (e.g. as a result of the proton-proton interactions, \cite{neronov07}), they would not be able to cool to  energies below $\sim 10$~MeV, which can explain the deficiency of the IC emission at the energies below the $\sim 5$~keV break energy.

Finally, the shape of electron spectrum below 10~MeV can be also affected by adiabatic cooling. Equating the adiabatic cooling rate 
$dE_e/dt=(E_e/R)dR/dt=nVE_e/(2D)$,
where $R\sim D^{n/2}$ is the radius of the synchrotron/IC emitting bubble and $V$ is the escape velocity, to the IC cooling rate, $dE_e/dt=(4/3)\sigma_TcU_{\rm ph}(E_e/m_ec^2)^2$, where $U_{\rm ph}=L_\star/4\pi D^2c$ is the radiation energy density, one finds the adiabatic break energy
\begin{equation}
E_{\rm ad}\simeq 10\mbox{ MeV}\left[\frac{D}{10^{13}\mbox{ cm}}\right]\left[\frac{V}{10^7\mbox{ cm/s}}\right]\left[\frac{L_\star}{10^{38}\mbox{ erg/s}}\right]^{-1}\mbox{ eV}
\end{equation}
If the emission is produced at the distances $D\sim 10^{13}$~cm, the observation of the cooling break at the energy $\epsilon\simeq 5$~keV is consistent with the assumption that the break is produced by the influence of the adiabatic losses if the escape velocity is $V\sim 10^7$~cm/s, which is 
in agreement with the estimates of the velocity of the stellar wind in the considered range of distances \citep{waters87}. 

The results of numerical modeling of the broad band spectrum of the source within the IC scenario of the X-ray emission are shown in Fig. \ref{fig:IC}. We assume that the high energy electrons are injected in the synchrotron/IC emitting bubble, which escapes with a speed equal to the speed of the stellar wind. The spectrum of the high-energy electrons, shown by the thick solid line in the upper panel of  Fig. \ref{fig:IC}, is formed as  a result of cooling of electrons due to the IC, synchrotron, bremsstrahlung and Coulomb energy losses during their escape from the system.
The injection of electrons is assumed to happen at a distance $D_0=3\times 10^{12}$~cm.
The injection spectrum of the electrons, shown by the dashed line in the upper panel of the figure is assumed to be a power law with spectral index $\Gamma_{\rm inj}=2.2$ \citep{kirk00}.
The magnetic field is assumed to be equal to $B_0=0.1$~G at the initial injection distance.
In the calculations presented in Fig. \ref{fig:IC} we have used the angle-averaged IC cross-section to calculate the IC energy loss and emission spectrum. The account of the anisotropy of the IC emission is important for the accurate calculations of the source light curve (see \cite{cher99, khangulyan08}), but requires a detailed 3D model and is not addressed here for simplicity. The cooled spectrum of electrons is shown on the top panel of Fig. \ref{fig:IC} with a solid line. Coulomb losses affect the electron spectrum at the lower end, leading to the formation of the first bump. The second bump arises due to the decrease of the cross-section after the transition from Thomson to the Klein-Nishina regime, which results in spectral hardening above $10^{11}$eV, and softening above $10^{12}$eV because of the dominance of the synchrotron losses at those energies.

\begin{figure}
\includegraphics[width=\linewidth]{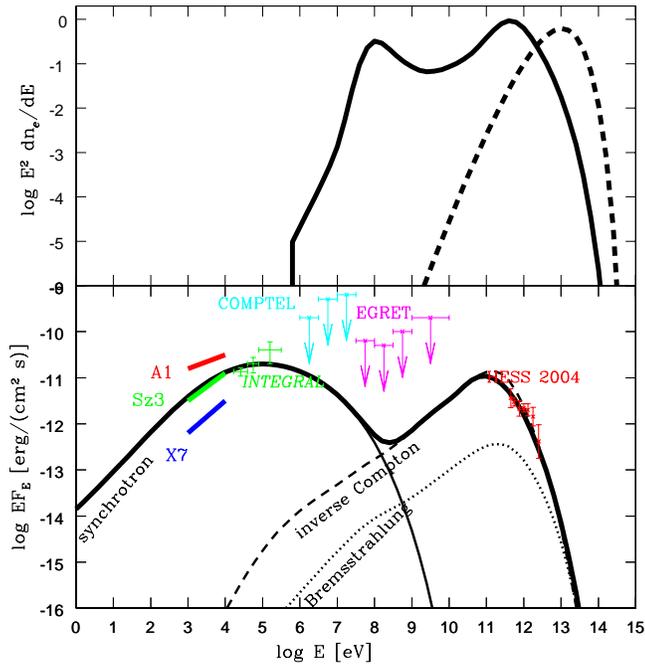}
\caption{Spectral energy distribution of \psrb\  in the model with electron injection at very high energies. Notations are the same as in Fig. \ref{fig:IC}.} 
\label{fig:synch}
\end{figure}

An alternative model, in which the observed X-rays are produced via the synchrotron mechanism is illustrated in Fig. \ref{fig:synch}.   For the calculation we have used the same numerical code as for the calculation shown in Fig. \ref{fig:IC}. The main differences between the two cases are different injection spectra and the assumed initial values of the magnetic field. The magnetic field at the initial distance $D=3\times 10^{12}$~cm is assumed to be $B_0=1$~G.  The electron injection spectrum, shown by the dashed line in the upper panel of Fig. \ref{fig:synch} is assumed to be sharply peaked at  energies about $10^{13}$~eV. If the typical initial electron injection energies are above 10~TeV,  the shape of the cooled electron spectrum at the energies below 10~TeV is completely determined by the cooling effects and is almost independent of the details of the initial electron injection spectrum.  

The low-energy cut-off in the cooled electron spectrum at  $E\sim 10^8$~eV arises because  at  energies below $10^8$~eV the radiative cooling time of electrons becomes longer than the escape time from the system. The spectrum of electrons is hard in the 10~GeV - 1~TeV energy interval, in which the electron energy losses are dominated by  IC scattering proceeding in the Klein-Nishina regime.  A sharp break in the spectrum at  TeV energies is related to the fact that the cooling rate of the multi-TeV electrons is determined by the synchrotron, rather than IC energy losses.  The break energy can be determined by equating the synchrotron loss time 
\begin{eqnarray}
\label{ts}
t_{S} = \frac{6\pi m_e^2c^3}{\sigma_T B^2 E_e} \simeq 4\times 10^2 \left[\frac{1G}{B}\right]^2\left[\frac{1\mbox{ TeV}}{E_e}\right] \mbox s
\end{eqnarray}
to the IC loss time in the Klein-Nishina regime,
\begin{eqnarray}
\label{tkn}
t_{KN}&\simeq& \frac{16E_eD^2\hbar}{\sigma_T (m_eckT_\star R_\star)^2}\ln^{-1}\frac{0.55E_ekT_\star}{m_e^2c^4} \nonumber\\
&\simeq& 10^3\left[\frac{E_e}{1 \mbox{TeV}}\right] \left[\frac{D}{10^{13} \mbox{cm}}\right]\mbox s 
\end{eqnarray}
This gives
\begin{equation}
E_{\rm br}=0.5\left[\frac{B}{1\mbox{ G}}\right]\left[\frac{D}{10^{13}\mbox{ cm}}\right]^{-1/2}\mbox{ TeV}
\end{equation}
The synchrotron emission produced by electrons with the energy $E_{\rm br}$ is emitted at the energy
\beq
\label{es}
\epsilon_{S}=\frac{e\hbar B E_e^2 }{m_e^3c^5} \simeq 10 \left[\frac{B}{1\mbox{ G}}\right] \left[\frac{E_e}{0.5\mbox{ TeV}}\right]^2 \mbox{keV}
\eeq
which is close to the observed break energy $\epsilon\simeq 5$~keV in the {\it Suzaku} spectrum.

The two different mechanisms of the X-ray emission (IC or synchrotron) could also be distinguished by the difference in the variability properties of the X-ray signal expected in the two models.  In principle, much faster variability is expected in the synchrotron model of X-ray emission. Within this model, the typical variability time scale is set up by the synchrotron and IC cooling times of the multi-TeV electrons. This time scale could be as short as $\le 1$~ks, if the magnetic field in the emission region is strong enough. On the other hand, within the IC model of X-ray emission, the characteristic variability time scales are set up by the rate of the Coulomb and/or IC and/or adiabatic energy losses of  $\sim 10$~MeV electrons, which are normally much longer than $\sim 1$~ks. The detection of the  variability in the X11 observation of \xmm\ at the time scale of $\sim 3$~ks thus provides an argument in favor of the synchrotron model of the X-ray emission. However,  fast variability in the X11 observation is detected at a $3\sigma$ level with the help of the structure function analysis. Further observations are necessary to firmly establish the presence/absence of the fast variability of the X-ray emission.

To summarize, we find that the X-ray data alone do not allow us to distinguish between the synchrotron and IC origin of the X-ray emission from the source.  The observation of the hardening of the spectrum below $\Gamma=1.5$ during the pre-periastron disk passage gives an important clue about the X-ray emission mechanism. However, the origin of the observed spectral hardening can be clarified  with the help of the simultaneous TeV observations. If the observed X rays have an IC origin, then the observed hardening during the drop of the flux is primarly connected to the hardening of electron spectrum below $\sim 10$~MeV (due to the Coulomb losses, or because of the escape of electrons from the emission region), so that no tight correlation between the X-ray spectral evolution and the TeV energy band emission is expected. On the other hand,  in the case of synchrotron origin of the observed X-rays, the spectral hardening can be produced if  the electron cooling is dominated by the IC energy loss in the Klein-Nishina regime. This implies that the IC flux from the system in the very-high-energy band at the moment of the spectral hardening should dominate over the X-ray flux. GeV data can give us another possibility to distinguish between the models, as in this region the predicted flux value is very different in IC and synchrotron model. Hopefully  \textit{Fermi} Gamma-ray telescope  observations of the next \psrb\ periastron passage will give us a clue to the distinguish between the models.

\section{Acknowledgments}
Authors are grateful to the unknown referee for valuable comments.

 \label{lastpage}

\end{document}